\documentclass[pra,amsmath,amssymb,twocolumn,superscriptaddress]{revtex4}
\usepackage{graphicx,epsfig}
\usepackage{amssymb,times}
\usepackage{amsmath,amsfonts,bm}
\usepackage{grffile}
\usepackage{dcolumn}
\graphicspath{{../plots_new/}}

\def\beq{\begin{equation}}
\def\eeq{\end{equation}}

\newcommand{\ket}[1]{| #1 \rangle}
\newcommand{\bra}[1]{\langle #1 |}
\newcommand{\aver}[1]{\langle #1 \rangle}
\newcommand{\bop}{\hat{b}} 
\newcommand{\bdop}{\hat{b}^{\dagger}} 
\newcommand{\rmo}{\hat{\rho}}
\newcommand{\rv}{\boldsymbol{\rho}}

\newcommand{\sutd}{Singapore University of Technology and Design, 20 Dover Drive, 138682 Singapore}
\newcommand{\unige}{D\'epartement de Physique Th\'eorique, Universit\'e de Gen\`eve, CH-1211 Gen\`eve, Switzerland.}
\newcommand{\ubc}{Department of Physics and Astronomy, University of British Columbia, Vancouver V6T 1Z1, Canada.} 
\newcommand{\cf}{Coll\`ege de France, 11 place Marcelin Berthelot, 75005 Paris, France.} 
\newcommand{\ep}{Centre de Physique Th\'eorique, Ecole Polytechnique, CNRS, 91128 Palaiseau Cedex, France.}
\newcommand{\dpmc}{DPMC-MaNEP,  Universit\'e de Gen\`eve, CH-1211 Gen\`eve, Switzerland.}

\begin{document}

\title{Dissipative quantum systems: from two to many atoms}

 \author{Dario Poletti}
 \affiliation{\sutd}
 \author{Jean-S\'ebastien Bernier}
 \affiliation{\ubc}
 \author{Antoine Georges}
 \affiliation{\cf}
 \affiliation{\ep}
 \affiliation{\dpmc}
 \author{Corinna Kollath}
 \affiliation{\unige}   

\begin{abstract}
We study the dynamics of bosonic atoms in a double well potential under the influence of dissipation. 
The main effect of dissipation is to destroy 
quantum coherence and to drive the system towards a 
unique steady state. We study how the atom-atom interaction 
affects the decoherence process. We use a systematic approach considering different atomic densities. 
We show that, for two atoms, the interaction
already strongly suppresses decoherence: 
a phenomenon we refer to as ``interaction impeded decoherence''.
For many atoms, thanks to the increased complexity of the system, the nature of the decoherence process 
is dramatically altered 
giving rise to an algebraic instead of exponential decay.   
\end{abstract}

\date{\today}

\maketitle
\section{Introduction}

Every device, may it be classical or quantum, operates surrounded by an environment which 
affects to a certain degree its properties. 
Understanding the interplay between the different physical processes 
at work and the dissipative processes induced by the environment is thus of paramount importance.
This is of particular 
interest for the development of technologies based on quantum mechanics 
as the environment can, for example, destroy the quantum coherence between different parts of the device 
and thus completely spoil its functionalities. 
Various quantum systems 
affected by dissipative processes have been studied in 
solid-state physics~\cite{LeggettZwerger1987,SchoenZaikin1990,BreuerPetruccione2002,Weissbook2008}, 
atomic and molecular physics~\cite{CohenTannoudjiGrynberg1998} and 
quantum optics~\cite{GardinerZollerBook,CarmichaelBook}. 
However, at the moment, 
a complete understanding of the physics of strongly correlated quantum many-body systems coupled to 
an environment is lacking.
In recent years, due to rapid technological progress, cold atom experimentalists have been 
able to build quantum emulators, systems
mimicking with a high degree of precision modeled Hamiltonians.
For this reason, these cold atoms have become an ideal test-bed 
to study quantum dynamical systems. 
In fact, the theoretical study of ultracold atoms paved the way to the 
first understanding of the interplay between interactions and dissipation.
Some notable works are (i) the discovery of Zeno-like dynamics due to local losses of atoms in a 
double well geometry~\cite{KhodorkovskyVardi2008,ShchesnovishKonotop2010,ShchesnovichMogilevtsev2010} 
or in optical lattices~\cite{BarmettlerKollath2011}, (ii) the existence of augmented stability against dissipation 
for weakly~\cite{WitthautWimberger2011} and strongly~\cite{PichlerZoller2010, PolettiKollath2012} 
interacting systems and (iii) the emergence of a dynamical phase transition between 
the condensed and thermal steady states~\cite{TomadinZoller2011}. 
Also of great importance are works detailing methods to engineer 
dissipative processes generating dark states with precise 
quantum properties~\cite{SyassenDuerr2008,DiehlZoller2008,Garcia-RipollCirac2009,TomadinZoller2011,
KantianDaley2009,VerstraeteCirac2009}.  

In this work we will study a quantum well in a systematic way by investigating different boson densities.
The system under study is a collection of ultracold atoms in a double well potential under 
the influence of a dissipative process. 
As a warm-up, we first consider a double well loaded with only
two atoms. This exercise will help us gain a deeper analytical understanding of the combined
effect of interaction and dissipation, and show that
this interplay leads to a marked slowing down of the decoherence process. 
We then study a system with many atoms and demonstrate that the increased complexity of the system
results in the emergence of a regime where coherence decays algebraically. 
Hence, for large atom numbers and interaction strength the system is particularly resistant to
dissipation.

This paper is organized as follows. In Section \ref{sec:model} we introduce the model and the various relevant parameters. 
In Section \ref{sec:two} we first study in detail the case of two atoms in a double well before investigating in \ref{sec:many} the dynamics for the same double well loaded this time with a large number of atoms. 
In Section \ref{sec:conc} we draw our conclusions.

\section{The model} \label{sec:model}  

The properties of an ensemble of bosonic atoms in a double well can be described, in the single band approximation, 
by the Hamiltonian 
\begin{equation}
 \hat{H}=-J\left(\bdop_1\bop_2 + \bdop_2\bop_1\right) + \frac U 2 \sum_{j=1,2} \hat{n}_j(\hat{n}_j-1). \label{eq:Hamiltonian} 
\end{equation}
Here $\bop_j$ ($\bdop_j$) is a bosonic operator
annihilating (creating) an atom at site $j$ and $\hat{n}_j=\bdop_j\bop_j$ counts the number of atoms at site $j$. 
Dealing with a double well, $j$ can only take the values $1,2$. The parameter $J$ is the hopping amplitude,
and $U$ the interaction strength.
As the ratio $U/J$ increases, the phase coherence between the two sites, 
$C=\aver{\bdop_1\bop_2+\bdop_2\bop_1}$, decreases.
To represent a state vector, we use the Fock basis $\ket{l,m}$ 
where $l$ and $m$ label the number of atoms on sites $1$ and $2$, respectively. 

%
As this ultracold gas of bosons is under the influence of a dissipative process, 
the system can be described by a density matrix $\rmo$ which evolves following the master equation 
\begin{equation}
\frac{d}{dt} \rmo = -\frac{i}{\hbar}\left[\hat{H},\rmo\right] +\gamma\mathcal{D}(\rmo) \label{eq:master}
\end{equation}
where $D(\rmo)$ is the dissipator modeled by  
\begin{equation}
\mathcal{D}(\rmo)=-\sum_{j=1,2} \frac{1} 2 \left[ \hat{n}_j,\left[\hat{n}_j,\rmo\right]\right] \label{eq:dissipator}  
\end{equation}
with coupling strength $\gamma$. It has been shown in~\cite{GerbierCastin2010,PichlerZoller2010} that this dissipator, written in Lindblad form~\cite{Lindblad1976,GoriniSudarshan1976}, 
can be experimentally realized, for example, in the context of ultracold atoms exchanging energy with a red-detuned optical lattice. 
Let us note that the considered dissipator leaves the Fock states invariant. 
Here we study the situation where the system is first prepared in the ground state of the Hamiltonian. At time $t=0$ the dissipation is switched on and the evolution of the system under the influence of both the Hamiltonian and the dissipator is investigated.

\section{Two atoms}   \label{sec:two}
Having established 
the model and main notations in the previous section we now attempt to gain a deeper 
understanding of the physics of this system. 
The best starting point is to study a double well loaded with two atoms as this limit 
can be solved analytically.
We show here that the system converges towards a unique steady state and 
that the relaxation process slows down as the interaction strength increases.
 
As a reminder, we first provide here a brief summary 
of the physics of the Hamiltonian system without dissipation. 
The dimension of the Hilbert space is three and we choose as a basis the 
states $\ket{2,0}$, $\ket{1,1}$, $\ket{0,2}$ in this order.   
The Hamiltonian in this basis is thus given by 
\begin{equation}
 \hat{H}=\begin{bmatrix}
  U & -\sqrt{2}J & 0 \\ 
  -\sqrt{2}J & 0 & -\sqrt{2}J \\ 
  0 & -\sqrt{2}J & U 
 \end{bmatrix}.
\end{equation}

The ground state of this Hamiltonian, which we choose as the initial state for the time evolution, 
has an energy $E_G=\frac 1 2 \left(U-\sqrt{U^2+16J^2}\right)$ and is given 
by $\psi_G=\alpha(1, \left(U+\sqrt{U^2+16J^2}\right)/\sqrt{8}J,1)$, 
where $\alpha$ is the normalization constant $\alpha=1/\sqrt{2+\left(U+\sqrt{U^2+16J^2}\right)^2/8J^2}$. 
Note that for large $U/J$ the ground state wavefunction tends towards $(\sqrt{2}J/U,1-2J^2/U^2,\sqrt{2}J/U)$ 
with energy $E_G\approx-\frac{4J^2} U $.          

With these results in mind, we can now address the following question: how does the system evolve under 
the effect of the dissipation considered? Using (\ref{eq:master}), we compute the time evolution of the nine 
elements of the density matrix $\rmo$. The evolution of the six complex independent 
elements \cite{Hermitian} is given by   
\begin{eqnarray}
\hbar\dot \rho_{2,2}&=&i\sqrt{2}J\left(\rho_{1,2}-\rho_{2,1}\right) \label{eq:eqrho}  \\   
\hbar\dot \rho_{1,1}&=&i\sqrt{2}J\left(\rho_{2,1}-\rho_{1,2} + \rho_{0,1}-\rho_{1,0} \right) \nonumber\\   
\hbar\dot \rho_{0,0}&=&i\sqrt{2}J\left(\rho_{1,0}-\rho_{0,1}\right) \nonumber\\   
\hbar\dot \rho_{2,1}&=&i\sqrt{2}J\left(\rho_{1,1}-\rho_{2,2} -\rho_{2,0} \right) -iU\rho_{2,1}-\hbar\gamma\rho_{2,1}\nonumber\\ 
\hbar\dot \rho_{2,0}&=&i\sqrt{2}J\left(\rho_{1,0}-\rho_{2,1}  \right) -4\hbar\gamma\rho_{2,0}\nonumber\\ 
\hbar\dot \rho_{1,0}&=&i\sqrt{2}J\left(\rho_{2,0}+\rho_{0,0} -\rho_{1,1} \right) +iU\rho_{1,0}-\hbar\gamma\rho_{1,0}. \nonumber 
\end{eqnarray}

Given that we prepare the system in the ground state of the Hamiltonian, the initial state is symmetric 
under the exchange of the first and third element. For this reason we 
only focus on the symmetric subspace of the density matrix evolution. 
Eq.~(\ref{eq:eqrho}) is a linear equation and its symmetric eigenvalues $\lambda$ are 
the roots of the polynomial $p(\lambda)=48 (\hbar\gamma)^2 J^2 + [4 (\hbar\gamma)^3 + 64 \hbar\gamma J^2 
+ 4 \hbar\gamma U^2] \lambda + [9 (\hbar\gamma)^2 + 16 J^2 + U^2] \lambda^2 + 6 \hbar\gamma \lambda^3 + \lambda^4 $, 
and the steady state, 
with eigenvalue $\lambda_S=0$, corresponds to the density matrix 
\begin{equation}
\rmo_{S}= \frac 1 3\begin{bmatrix} 
  1 & 0 & 0 \\ 
0 & 1 & 0 \\ 
0 & 0 & 1 
 \end{bmatrix} \label{eq:rhos}  
\end{equation} 
which is the completely mixed state (or the highest entropy state). 
The roots of $p(\lambda)$ are in general complex numbers with $Re(\lambda)<0$. 

\subsection{Dynamical transitions}  

Studying the roots of $p(\lambda)$, we identify dynamical transitions which arise when  
tuning $U/J$ and $\hbar\gamma/J$ transforms a real into a complex eigenvalue (or the converse). 
Dynamically this implies that a simple exponential decay, the hallmark of real eigenvalues, can
be transformed into an exponential decay accompanied by oscillations (due to the imaginary part). The oscillations have their origin in the Hamiltonian evolution. The presence of dynamical transitions becomes particularly interesting if one changes the value of 
the different parameters, for example $U$ or $\gamma$, in time, allowing for the modulation of the 
dynamics from an overdamped to an underdamped regime. 
\begin{figure}[!ht]
 \includegraphics[width=0.9\linewidth]{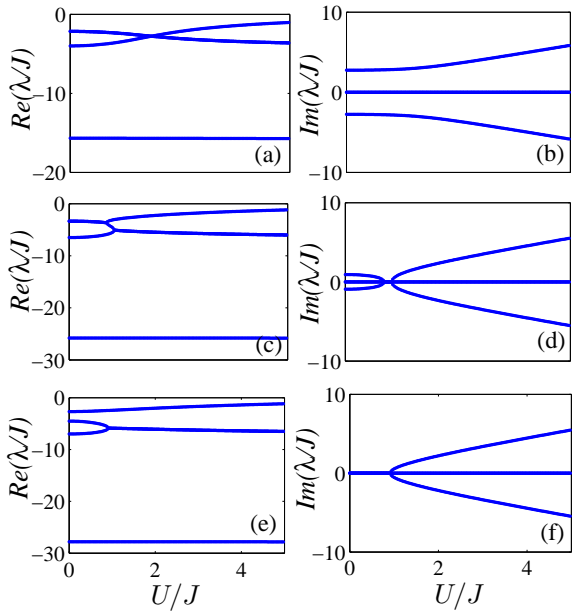}
   \caption{Real (a,c,e) and imaginary (b,d,f) part of the eigenvalues $\lambda$ (divided by J) 
            versus the interaction $U/J$ for $\hbar\gamma/J=4$ (a,b), $\hbar\gamma/J=6.5$ (c,d) 
            and $\hbar\gamma/J=7$ (e,f).} \label{fig:Eigs}  
\end{figure}
The various dynamical scenarios are depicted in Fig.~\ref{fig:Eigs} for different values of $\hbar\gamma/J$. 
For $\hbar\gamma/J\lesssim 6.3$, as show in Fig.~\ref{fig:Eigs}(a,b), two eigenvalues 
are real and two are complex conjugates. For $6.3 \lesssim \hbar\gamma/J\lesssim 6.8$, Fig.~\ref{fig:Eigs}(c,d), 
there are three regimes as a function of $U/J$: in the first regime, there are two complex and two real $\lambda$, 
in the second regime all the $\lambda$ are real, and in the last regime again two eigenvalues 
are real and two are complex. For larger values of the dissipative coupling, $\hbar\gamma/J\gtrsim 6.8$, 
there are two regimes: for small $U/J$, the decays are only exponential, while for large $U/J$ two eigenvalues 
are real and two are complex (see Fig.~\ref{fig:Eigs}(e,f)). 

For any value of $\hbar\gamma/J$, if $U/J$ is large enough the imaginary part of two of the eigenvalues 
will be proportional to $\pm U/J$ indicating the strong influence of the Hamiltonian 
part of the dynamics. 

\begin{figure}[!ht]
 \includegraphics[width=0.9\linewidth]{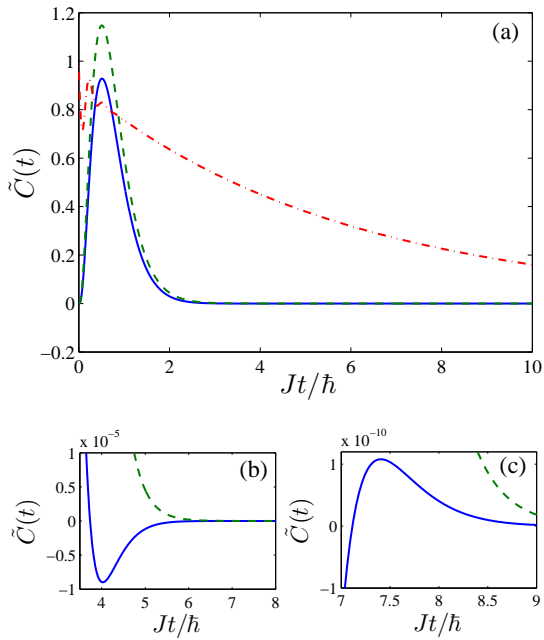}
   \caption{Absolute value of the coherence versus time in a log-log plot for $\hbar\gamma/J=6.5$ and for interaction $U/J=0.1$ (blue solid line), 
            $U/J=1$ (green dashed line), $U/J=20$ (red dot-dashed line). The inset shows the coherence versus time in a linear plot to highlight the oscillating behavior for $U/J=20$.} \label{fig:dynbeha}  
\end{figure}

The change in dynamics is exemplified in Fig.~\ref{fig:dynbeha} where $\hbar\gamma/J=6.5$. 
In this figure, we plot the absolute value of the coherence versus time in a log-log form which highlight the change of sign of the coherence for $U/J=0.1$ (blue line). We observe that, 
as expected, for both $U/J=0.01$ (blue solid line) and $U/J=20$ (red dot-dashed line) the evolution of the coherence 
is oscillatory while for $U/J=1$ (green dashed line) there are no oscillations but only a sum of exponential decays. 
The inset highlights the oscillatory behavior for $U/J=20$. 
%
 
\subsection{Quantum Zeno effect versus interaction impeded decoherence} 

The long time dynamics of the system are governed by its slowest decaying state, those whose 
corresponding eigenvalues have the smallest (in modulus) real part. For this small system with 
only five symmetric states, the relevant ones are the steady state $\rmo_S$ and the next slowest 
decaying state $\rmo_E$ with corresponding $\lambda_E$. While the steady state is independent 
of $U/J$, $\rmo_E$ and its decay rate can change considerably with $U/J$. 
To gain a deeper understanding of how the interaction affects the dynamics, we first present 
the non-interacting case $U=0$ and then cover the large interaction limit where $U/J\gg 1$ 
and $U/\hbar\gamma \gg 1$.   

\subsubsection{Quantum Zeno effect} 
For the non-interacting case, $U=0$, the equation for the eigenvalues $p(\lambda)$ takes the simpler form 
$p(\lambda)_{U=0}=(\hbar\gamma+\lambda)\left\{48 \hbar\gamma J^2 + [4 (\hbar\gamma)^2 + 16 J^2] \lambda + 5 \hbar\gamma \lambda^2 + \lambda^3\right\}$. 
The slowest decaying eigenvalue at large $\hbar\gamma/J$ is, in this regime, 
\begin{equation}
\lambda_E=\lambda_{QZ}=-12\frac{J^2}{\hbar\gamma}. \label{eq:lQZ} 
\end{equation} 
This is a clear manifestation of the Quantum Zeno effect: increasing the dissipative coupling $\gamma$
suppresses the decay rate as the continuous density measurement prevents the system from evolving. 
An experimental observation of this effect was reported in \cite{ItanoWineland1990}. 

\subsubsection{Interaction impeded decoherence}  
When the interaction is the dominant energy scale (for large $U/J$ and large $U/\hbar\gamma$), 
we analytically find that 
\begin{equation}
\lambda_E=\lambda_I=-12\frac{\hbar\gamma J^2}{U^2}. \label{eq:lI}
\end{equation} 
Similarly to the Quantum Zeno regime, here too the decay is strongly suppressed. However, 
as the physical mechanism behind this slowing down is different from the Quantum Zeno case,
the dependence of $\lambda_I$ on the system parameters is distinct from $\lambda_{QZ}$.
Remarkably, in this regime, the presence of interactions renders the system more robust against dissipation, 
as it increases the size of the energy gap that needs to be overcome to populate higher states. 
As reaching the steady $\rho_S$ requires a large amount of energy, of the order of $U$, the dissipative process, heating up the system, will need a longer time to provide this energy at large $U$. 
Consequently, we refer to this phenomenon as ``interaction impeded decoherence''. 
%

\begin{figure}[!ht]
 \includegraphics[width=0.9\columnwidth]{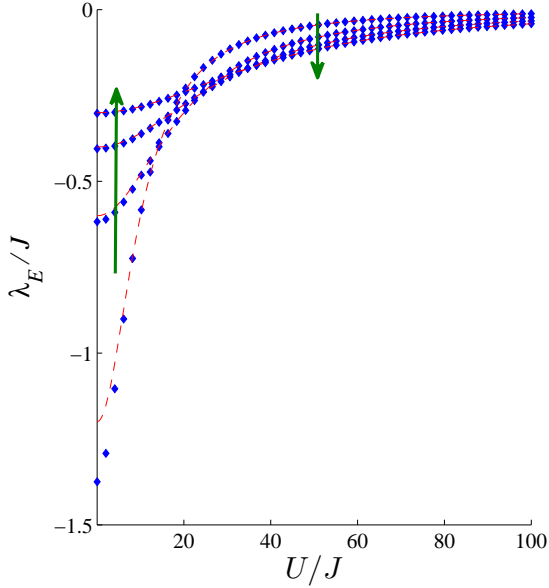}
   \caption{(a) Slowest decaying even eigenvalue $\lambda_E/J$ versus interaction $U/J$ (blue diamonds) for 
            different $\hbar\gamma/J=10,20,30,40$. The blue diamonds indicate the exact numerical values 
            while the red-dashed line represents the approximate solution from (\ref{eq:lQZI}). 
            (b) $\lambda_E/J$ times $\hbar\gamma/J$ versus interaction $U/J$. 
            (c) $\lambda_E/J$ divided by $\hbar\gamma/J$ versus interaction $U/J$. 
            The arrows indicate increasing $\gamma$ values.} \label{fig:EigCrossInter}  
\end{figure}

The cross-over between the two regimes is clearly illustrated in Fig.~\ref{fig:EigCrossInter}(a) where, 
for small interactions, $\lambda$ decreases as $\gamma$ increases, while, for large interactions, 
$\lambda$ and $\gamma$ increase together as indicated by the green arrows. 

For higher clarity we have also shown in Fig.~\ref{fig:EigCrossInter}(b,c) the value of, respectively, $(\lambda_E/J)\times(\hbar\gamma/J)$ and $(\lambda_E/J)/(\hbar\gamma/J)$. 
The collapse of the curve in Fig.~\ref{fig:EigCrossInter}(b) confirms the emergence of a Quantum Zeno behavior while the collapse of the curve in Fig.~\ref{fig:EigCrossInter}(c) 
signals the emergence of the ``interaction impeded coherence'' behavior.  

\subsection{Large dissipation or large interaction limit}  

We gain deeper insight into the large interaction $U\gg J$ and/or large dissipation regime $\hbar\gamma\gg J$ 
via an adiabatic elimination technique~\cite{SaitoKayanuma2002, PolettiKollath2012}. 
The main idea behind this method is that at long times the dynamics is dominated by the diagonal elements 
of $\rmo$ which dictate the full quantum dynamics. In fact, their evolution relates to the hopping 
parameter $J$, the smallest parameter in the system, and connects the diagonal and the 
off-diagonal elements. However, the off-diagonal elements evolve very quickly, with a dephasing due to the interaction $U$ 
and/or a fast exponential decay due to $\gamma$. Hence, the evolution of the off-diagonal terms can be integrated 
out and we are left with a coarse-grained evolution for the whole system. 

More precisely let us take, for example, the evolution of $\rho_{0,1}$ from (\ref{eq:eqrho}). This is given by  
\begin{eqnarray}
 \rho_{0,1}(t) &\approx&  e^{-(i\frac{U}{\hbar}+\gamma)t}\rho_{0,1}(0) \label{eq:ae1} \\ 
&+&i\frac{\sqrt{2}J}{\hbar} e^{-(i\frac{U}{\hbar}+\gamma)t} \int_0^t  e^{(i\frac{U}{\hbar}+\gamma)\tau} \left(\rho_{1,1}-\rho_{0,0}-\rho_{0,2}\right) d\tau. \nonumber  
\end{eqnarray}   
Here it should be noted that $\rho_{0,2}$, being far from the diagonal, is much smaller and decays even faster. 
Also as $U$ and $\gamma$ are large, we can integrate (\ref{eq:ae1}) by parts to obtain \cite{SaitoKayanuma2002,PolettiKollath2012} 
\begin{eqnarray}
 \rho_{0,1}(t) &\approx& \sqrt{2}J \frac{\rho_{1,1}-\rho_{0,0}}{U-i\hbar\gamma}  \nonumber \\  
&\approx& \sqrt{2}J \frac{U+i\hbar\gamma}{U^2+(\hbar\gamma)^2}\left(\rho_{1,1}-\rho_{0,0}\right). \label{eq:ae2}
\end{eqnarray}
Likewise we have   
\begin{eqnarray}
 \rho_{1,2}(t) &\approx&  \sqrt{2}J \; \frac{U-i\hbar\gamma}{U^2+(\hbar\gamma)^2} \left(\rho_{2,2}-\rho_{1,1}\right) \nonumber \\             
 \rho_{0,2}&\approx&0. \label{eq:ae3}     
\end{eqnarray}
We now inject the information from (\ref{eq:ae2}) and (\ref{eq:ae3}) into the time evolution of the 
diagonal elements and, as anticipated earlier, the diagonal values of $\rmo$ are determined by three independent equations. This results in 
\begin{subequations}
\begin{eqnarray}
 &&\frac d {dt} \left(\rho_{0,0}+\rho_{1,1}+\rho_{2,2}\right)  = 0 \label{eq:ae4a} \\ 
 &&\frac d {dt} \left(\rho_{0,0}-\rho_{2,2}\right) \approx \frac{\lambda_O}{\hbar} \left(\rho_{0,0}-\rho_{2,2}\right) \label{eq:ae4b} \\ 
 &&\frac d {dt} \left(\rho_{0,0}-2\rho_{1,1}+\rho_{2,2}\right) \approx \frac{\lambda_E }{\hbar}
\left(\rho_{0,0}-2\rho_{1,1}+\rho_{2,2}\right). \nonumber \\ \label{eq:ae4c} 
\end{eqnarray}
\end{subequations}
Eq.~(\ref{eq:ae4a}) shows consistently that $\rmo_S$ is the steady state, (\ref{eq:ae4b}) refers to the 
slow decaying anti-symmetric solution, which decays at a rate $\lambda_O/\hbar=-4J^2\gamma/(U^2+(\hbar\gamma)^2)$ but 
is not relevant for our dynamics, and the last equation, (\ref{eq:ae4c}), represents the symmetric 
slowest decaying state with a rate  
\begin{equation}
\frac{\lambda_E}{\hbar}=-12\frac{ \gamma J^2}{U^2+(\hbar\gamma)^2}. \label{eq:lQZI}
\end{equation} 
From (\ref{eq:lQZI}), we derive both the Quantum Zeno behavior of (\ref{eq:lQZ}) and the 
``interaction impeded decoherence'' behavior (\ref{eq:lI}) in their respective regimes (small and large 
interactions limits). 
     
Using (\ref{eq:ae2}) and (\ref{eq:ae3}), we compute, at any given time, the coherence of the system knowing 
only the terms on the diagonal. In particular, the coherence of the slowest decaying state $\lambda_E$ is 
given by 
\begin{equation}   
C_E\propto\frac{JU}{U^2+(\hbar\gamma)^2}
\end{equation}     
which is proportional to $U$ when it is small compared to $\hbar\gamma$, and inversely 
proportional to $U$ when it is the dominant energy scale. Thus the large interaction strengths 
affect the coherence in two ways: $U$ lowers the coherence of the slowest decaying state, but at the same time 
reduces the decay rate $\rmo_E$.    

\section{Many atoms} \label{sec:many}

We now want to understand how the dynamics is altered by the presence of many atoms. 
As for the two atom case the interaction was found to play a crucial role in the dynamics, 
we can expect that the presence of many interacting atoms will affect the dynamics even 
more dramatically. We first present a perturbative study at short times highlighting the subtle 
role played by interactions, we then cover in detail the strongly interacting regime where coherence 
decays algebraically with time. 

\subsection{``Interaction impeded decoherence'': first seeds}  

To illustrate how interaction influences decoherence, we study perturbatively the time evolution of the coherence. 
We first show analytically that for a non-interacting system the coherence decays following 
a simple exponential. To prove this statement it is sufficient to show that the evolution of 
the operator $\hat{C}=\bdop_1\bop_2+\bdop_2\bop_1$ follows  
\begin{eqnarray}
 \frac{d}{dt}  \hat{C} &=& \frac i \hbar\left[\hat{H},\hat{C}\right] +\gamma\mathcal{D}(\hat{C}) \nonumber \\
 &=&  \gamma\mathcal{D}(\hat{C}) \nonumber \\ 
 &=&  -\gamma\hat{C}. \label{eq:masterfor C}
\end{eqnarray}
This expression is derived from the fact that $\mathcal{D}(\hat{C})=\hat{C}$ and that, for $U=0$, 
$\left[\hat{H},\hat{C}\right]=0$. In the non-interacting limit, the evolution of the coherence is 
then simply described by 
\begin{eqnarray}
C(t)_{U=0}=e^{-\gamma t}C(0).
\end{eqnarray}
For non-zero interaction strength, and in particular for large values of $U$, the decoherence follows 
a different law even for short times. To demonstrate this statement, we analyze the evolution of the density 
matrix $\rmo$ up to third order in $dt$
\begin{equation}     
\rmo(t+dt)=\rmo(t)+\dot\rmo(t)dt+\frac{\ddot\rmo(t)}{2}dt^2+\frac{\dddot\rmo(t)}{6}dt^3+...      
\end{equation}    
where each superimposed dot stands for a time derivative. 
As we use the ground state of the Hamiltonian as the initial state, we have 
that $\left[\hat{H},\rmo_G\right]=0$ and hence can write 
\begin{eqnarray}     
\dot\rmo_G&=&\mathcal{D}(\rmo_G) \label{eq:pert} \\ 
\ddot\rmo_G&=&\mathcal{D}\left(\mathcal{D}(\rmo_G)\right) -\frac i {\hbar} \left[\hat{H},\mathcal{D}(\rmo_G)\right]\nonumber \\ 
\dddot\rmo_G&=&\mathcal{D}\left(\mathcal{D}\left(\mathcal{D}(\rmo_G)\right) \right)  
-\frac i {\hbar} \mathcal{D}\left(\left[\hat{H},\mathcal{D}(\rmo_G)\right] \right)  \nonumber \\  
&-&  \frac i {\hbar} \left[\hat{H},\mathcal{D}\left(\mathcal{D}(\rmo_G)\right)\right] 
- \frac 1 {\hbar^2} \left[\hat{H},\left[\hat{H},\mathcal{D}(\rmo_G)\right]\right].   \nonumber   
\end{eqnarray}    
The coherence is only dependent on the real part of the off-diagonal terms of the 
time-evolved density matrix, hence, for $U/J\gg 1$, we derive  
\begin{equation}
C(dt)=\left(1-\gamma dt+\frac{\gamma^2} 2 dt^2 - \frac{\gamma^3-\gamma (U/\hbar)^2}{6} dt^3\right)C(0).
\end{equation} 
If $U=0$, this expression would be the Taylor expansion of an exponential decay. 
However, due to the finite interaction strength, the short time evolution is slowed down as 
$U$ opposes the effect of $\gamma$. This constitutes a first glimpse into the role played by 
interactions. In the following, we derive an expression for the decay of the coherence $C$ 
at large times and for a large number of atoms. This expression provides a much deeper understanding 
of the emerging physics.     

\subsection{Eigenvalues problem} 

\begin{figure}[!ht]
 \includegraphics[width=0.9\linewidth]{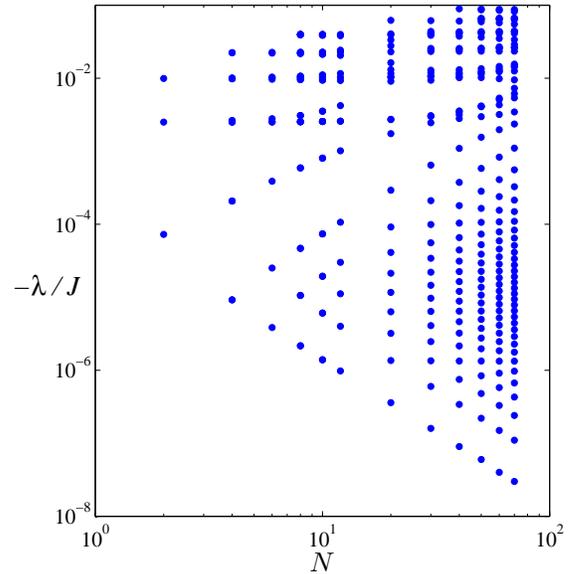}
   \caption{Real part of the eigenvalues $Re\left(\lambda_{\alpha}\right)$ as a function of the number of atoms $N$ for $\hbar\gamma/J=1$ and $U/J=20$. 
   The red-dashed line represent an algebraic fit that scales as $1/N^2$.}\label{fig:eigenvalues}
\end{figure}

\begin{figure}[!ht]
 \includegraphics[width=0.9\linewidth]{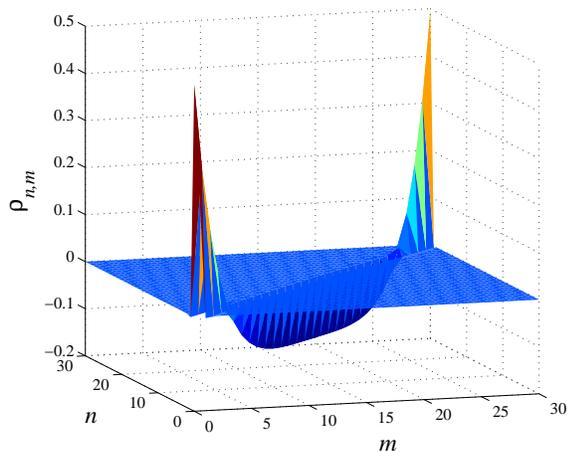}
   \caption{Typical plot of the density matrix $\rmo$ for the slowest decaying symmetric state. 
            This particular case is for $N=30$, $U/J=10$ and $\hbar\gamma/J=1$. }\label{fig:DMevensector}
\end{figure}

A very good first insight is provided by the study of the eigenvalues $\lambda$ 
of the evolution of the density matrix $\rmo$. For the general case with $N$ atoms, 
the size of the density matrix $\rmo$ is $(N+1)^2$.
The elements of the density matrix $\rho_{l,m}$ are such that 
$\rmo=\sum_{l,m}\rho_{l,m}\ket{l,N-l}\bra{m,N-m}$ where $N$ is the total number of atoms.  Rewriting the matrix $\rmo$ 
in a vector form $\rv$, we compute at each time $t$ the vector $\rv$. 
This vector is given by the sum over all $\rv_{\alpha}$, each associated with a 
complex decay constant $\lambda_{\alpha}$, 
\begin{equation}
\rv(t)=\sum_{\alpha} c_{\alpha}e^{\lambda_{\alpha}t/\hbar}\rv_{\alpha}
\end{equation}
here $c_{\alpha}$ is the weight of each $\rv_{\alpha}$ at time $t=0$. The real parts of the different $\lambda_{\alpha}$ 
are plotted in Fig.~\ref{fig:eigenvalues}. 
%
Two aspects of this figure are worth noting: (i) the slowest decaying state is gapped from 
all others (this gap depends on the number of particles as $1/N^2$) and (ii) for large $N$ an almost continuous band of slowly decaying states arises. The first point 
implies that past a certain critical time the decay will be fully exponential and solely due to the 
slowest decaying state $\rmo_E$ (note that the corresponding $\lambda_E$ decreases as $1/N^2$ as shown by the linear fit in Fig.~\ref{fig:eigenvalues}). This state, as expected, is mainly populated 
on the diagonal, see Fig.~\ref{fig:DMevensector}, and these diagonal elements follow a 
particular distribution. 
We observe that the matrix elements corresponding to large density imbalance 
between the left and right sites, for example $\rho_{0,0}$ or $\rho_{N,N}$, are much more populated than 
those with balanced configurations, for example $\rho_{N/2,N/2}$. For a large interaction strength, 
states with strong imbalance are highly energetic and thus populating them requires the dissipation to 
provide a tremendous amount of energy. Moreover, in order to populate the state $\rho_{n,n}$, with $n$ large, 
it is necessary to first populate state $\rho_{n-1,n-1}$ which in turns necessitates a long time.
Thus, the different configurations can be arranged into an energetical hierarchy.       

\begin{figure}[!ht]
 \includegraphics[width=0.9\columnwidth]{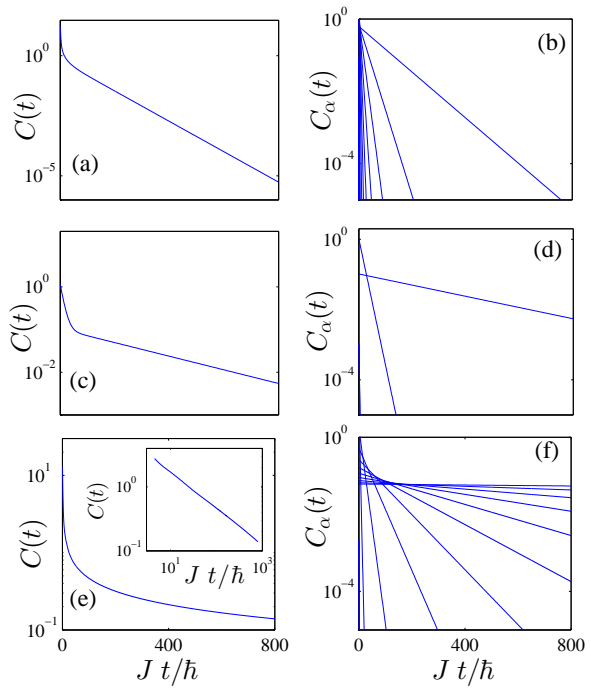}
   \caption{Left column: decay of the coherence $C(t)$ versus time for different values of $N$ and $U/J$. 
            Right column: decay of the weighted coherence of each eigenstate $C_{\alpha}(t)$ versus time. 
            The inset of Fig.~\ref{fig:decay_coherence}(e) presents a log-log of the coherence and highlight 
            the power-law regime. The parameters are $N=20$ and $U/J=2$ for (a) and (b), 
            $N=5$ and $U/J=20$ for (c) and (d), and $N=20$ and $U/J=20$ for (e) and (f), 
            while $\hbar\gamma/J=1$ for all the subplots.} \label{fig:decay_coherence} 
\end{figure}

The second aspect, the presence of a band of eigenvalues, paves the way to the emergence of a new decay regime.
This interesting behavior arises as similar exponential decays are summed over. The emergence of this regime 
is shown in Fig.~\ref{fig:decay_coherence} where coherence decay for various values of $U/J$ and $N$ is plotted. 
On the left panels ($a$,$c$,$e$), we show how the slower decaying states decay in time, while on the right panels ($b$,$d$,$f$) 
are plotted the weighted coherence of each eigenstate given by
\begin{equation}
C_{\alpha}(t)=c_{\alpha}e^{\lambda_{\alpha}t/\hbar}\;tr\left(\hat{C}\rho_{\alpha}\right).
\end{equation} 
For small $U/J$ or $N$ the exponential decay of the slowest decaying state $\rmo_E$ takes place earlier
as the interaction is not strong enough to impede decoherence 
(small interaction $U/J=2$ in Fig.~\ref{fig:decay_coherence}(a,b)) or there are not enough 
states to form a band (small atom number $N=5$ in Fig.~\ref{fig:decay_coherence}(c,d)). 
When both the interaction and atom number are sufficiently large, for example $U/J=20$ and 
$N=20$ as in Fig.~\ref{fig:decay_coherence}(e,f), a power-law decay regime emerges over a large time.
This regime is clearly shown on the log-log plot in the inset of Fig.~\ref{fig:decay_coherence}(e).        
%

\subsection{Emergence of an algebraic relaxation regime} 

To provide a better understanding of this emerging dynamical behavior and to pinpoint the algebraic 
nature of this regime, we first proceed with a coarse-graining of the evolution using adiabatic elimination 
and then the resulting equations are further approximated within a continuum limit. 
The end product of this procedure is a continuum classical diffusion equation 
with a non-uniform diffusion function. The solution of this diffusion equation is then used to 
demonstrate the existence of the algebraic regime \cite{PolettiKollath2012}.  

\subsubsection{Adiabatic elimination}   

As for the two-atom problem described in section III, we study the long time full quantum dynamics by
only considering the diagonal elements of $\rmo$. The adiabatic elimination, in the regime of large dissipation 
$\hbar\gamma/J$ and interaction $U/J$, thus reduces the size of the system to $(N+1)$. 
The system of equations describing the dynamics is derived by generalizing (\ref{eq:ae2}) to the  
limit $U/\hbar\gamma\gg 1$ to obtain 
\begin{eqnarray} 
   \rho_{n,n+1}(t) &\approx& \frac{J\sqrt{(n+1)(N-n)}}{U^2(N-2n-1)^2}\nonumber \\ 
   &&\times \left(U(N-2n-1)-i\hbar\gamma\right)\nonumber\\ 
   &&\times (\rho_{n+1,n+1}-\rho_{n,n}).  \label{eq:approxoffdiag}  
\end{eqnarray} 
This generalization is also applied to the master equation (\ref{eq:master}) and gives 
\begin{subequations}
\begin{eqnarray} 
 \partial_{\tau} \rho_{0,0} &=& \frac{N}{(N-1)^2} \left( \rho_{1,1} - \rho_{0,0} \right)    \label{eq:approxmodel1} \\ 
 \partial_{\tau} \rho_{n,n} &=& \frac{(n+1)(N-n)}{(N-2n-1)^2}\left(\rho_{n+1,n+1}-\rho_{n,n}\right)  \label{eq:approxmodel2} \\ 
&& + \frac{n(N-n+1)}{(N-2n+1)^2} \left(\rho_{n-1,n-1} - \rho_{n,n}\right) \nonumber \\ 
 \partial_{\tau} \rho_{N,N} &=& \frac{N}{(N-1)^2} \left( \rho_{N-1,N-1} - \rho_{N,N} \right) \label{eq:approxmodel3}            
\end{eqnarray}\label{eq:mastersimp}
\end{subequations}
for $n\ne \{0, N\}$, $N$ even and $\tau=2(J^2/U^2) \gamma t$. These last equations are used to accurately compute the dynamics 
and give access to the coherence of the system 
\begin{eqnarray} 
C &\approx& \sum_n\frac{2J(n+1)(N-n)}{U(N-2n-1)}(\rho_{n+1,n+1}-\rho_{n,n}). \nonumber \\ \label{eq:cohediscr}        
\end{eqnarray} 

\subsubsection{Continuum limit}

By performing the large $N$ limit, the discrete master equation (\ref{eq:mastersimp}) is mapped onto 
a classical diffusion equation. The configuration space, indexed by $n$ is mapped to the 
coordinate $x=n/N-1/2\in [-1/2,1/2]$ which is a continuous variable in the limit $N\rightarrow\infty$. 
The boundaries of this system then become $x= \pm 1/2$ corresponding to the strongly imbalanced 
configurations of the double well, $n=0$ and $n=N$, whereas the center, $x=0$, corresponds to the balanced 
configuration $n=N/2$. The diagonal elements of the density matrix $\rmo$ are connected to a 
continuum probability density distributions $p(x,\tau)$ via the relation $N\rho_{n,n}(\tau)=p(x,\tau)$. 
Within this limit, the normalization of the probability density distribution is given 
by $\int_{-1/2}^{1/2}p(x)\textrm{d} x =1$ which is equivalent to $\mathrm{tr}\,\hat{\rho}=1$. The 
initial state corresponds to $p(x,\tau=0)=\delta(x)$, and the steady state is then given by the uniform 
distribution $p(x,\tau=\infty)=1$ which represents the totally mixed state $\rmo_S$.  
To derive the differential equation for the probability density function $p(x,\tau)$, we rewrite 
(\ref{eq:approxmodel2}) as 
\begin{eqnarray} 
 \partial_{\tau} p(x,\tau) &=& \frac{(x+dx)(1-x)}{(1-2x-dx)^2}\left(p(x+dx,\tau)-p(x,\tau)\right)  \label{eq:continuum1} \nonumber \\ 
&& + \frac{x(1-x+dx)}{(1-2x+dx)^2} \left(p(x-dx,\tau)-p(x,\tau)\right) \nonumber \\ 
\end{eqnarray}
where $dx=1/N$. Expanding the right-hand side of (\ref{eq:continuum1}), to second order in $dx$, gives the diffusion equation
\begin{eqnarray}
\partial_{\tau_N} \; p(x,\tau_N)=\partial_x\left[D(x)\partial_x p(x,\tau_N)\right]  \label{eq:continuum}
\end{eqnarray}   
where $\tau_N=\tau/N^2$. The diffusion function $D(x)=\frac{1}{4x^2}-1$ varies substantially as a 
function of $x$: for the balanced configuration $x=0$, $D(x)$ diverges while for strongly imbalanced 
configurations $x\approx\pm 1/2$ it vanishes. Thus, if the initial state is near the origin, which is the case when 
the evolution begins from the ground state, we observe a very fast initial diffusion; when the 
probability density distribution approaches the boundaries the diffusion drastically slows down. 

For short times, such that $p(x,\tau_N)$ is not affected by the boundaries, the diffusion 
equation (\ref{eq:continuum}) is solved by using a scaling ansatz 
\begin{equation}
p(x,\tau_N)= \frac{1}{\tau_N^{\nu}}f(\xi) \label{eq:scaling}
\end{equation}
with $\xi=x/\tau_N^{\nu}$. A scaling solution exists for $\nu=1/4$ and $\tau_N\ll 1$ giving the differential equation for $f$ 
\begin{equation}
\xi f''+(\xi^4-2)f'+\xi^3f=0.
\end{equation}
This expression is solved analytically giving 
\begin{equation}
f(\xi)\propto \exp(-\xi^4/4).
\end{equation}
The diffusion for the probability density function $p(x,\tau_N)$ is thus, for short rescaled time $\tau_N\ll 1$, given by 
\begin{equation}
p(x,\tau_N)= \frac{\sqrt{2}}{\Gamma(1/4)} \frac{1}{\tau_N^{1/4}}\,\exp\left(-x^4/4\tau_N\right) \label{eq:proba}    
\end{equation}
where the constant $\Gamma(\zeta)$ is the gamma function with argument $\zeta$. 
Eq.~(\ref{eq:proba}) describes an anomalous diffusion process as $\aver{x^2} \propto \sqrt{\tau_N}$ which, since this formulation is valid for $\tau_N$ small, 
and as for normal diffusion $\aver{x^2} \propto \tau_N$, (\ref{eq:proba}) is an example of super-diffusion. 
Hence, the divergence in $D(x)$ at $x\approx 0$ leads to a highly accelerated initial diffusion. 

\begin{figure}[!ht] 
  \includegraphics[width=0.9\columnwidth]{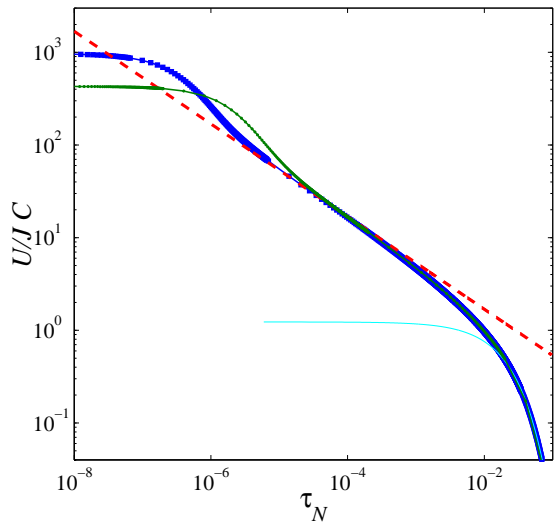}
  \caption{(color online) Coherence versus time for $N=60$, $U/J=20$ and $\hbar\gamma/J=1$ (blue squares), 
  for $N=50$, $U/J=10$ and $\hbar\gamma/J=0.1$ (green diamonds), for $N=8$, $U/J=20$ and $\hbar\gamma/J=5$ (light blue stars) and
  for $N=40$, $U/J=0.1$ and $\hbar\gamma/J=1$ (black continuous line). The red dashed line 
           represents the analytical result of (\ref{eq:coherence}). } \label{fig:coherencematch}         
\end{figure}

Regarding the coherence $C$, it is possible to obtain an analytical expression from the continuum limit of (\ref{eq:cohediscr}): 
\begin{eqnarray} 
\frac{C}{N}&=& \frac{J}{UN} \int_{-1/2}^{1/2}\frac{x^2-1/4}{x} \partial_x p(x,\tau_N) {\textrm dx}  \\   
&=& \frac{J}{UN} \frac{\Gamma(3/4)}{2\Gamma(1/4)} \frac{1}{\sqrt{\tau_N}}
=\frac{\Gamma(3/4)}{\sqrt{2}\Gamma(1/4)} \frac{1}{\sqrt{\gamma t}}. \label{eq:coherence}   
\end{eqnarray}  
This expression is in excellent agreement with the numerical simulations (see Fig.~\ref{fig:coherencematch}). 
Hence, we demonstrated here how the power-law regime emerges for a large number of atoms and strong 
interactions. Fig.~\ref{fig:coherencematch} also shows clearly that if the interaction is not strong enough (black continuous curve for $U/J=0.1$) the power-law regime does not 
emerge. It also shows that the algebraic region will increase when more atoms are in the system. 
In fact the power-law region is larger for large $N$, for example $N=60$ (blue squares), and it 
is smaller for fewer atoms, for example $N=8$ (light blue stars). 
From Fig.~\ref{fig:coherencematch}, we also notice that for smaller $\gamma$ the algebraic
region begins at later times and is thus reduced (see green diamonds curve for $N=50$, $U/J=10$ and $\hbar/J=0.1$). 
The fact that the time-evolution of the coherence is independent of the interaction strength
is particular to the coherence in this problem.  
\section{Conclusions} \label{sec:conc}

To summarize we have shown here how the interaction between atoms can affect decoherence in a quantum system. 
We analyzed a double well setup by first considering only a system of two atoms and by subsequently 
extending our study to a system of many atoms. For two atoms, we discovered that various dynamical 
behaviors emerge and depend on the ratios between $U/J$ and $U/\hbar\gamma$. We also showed that 
strong interactions suppress strongly the detrimental effects due to dissipation. We refer to this process as ``interaction impeded decoherence''. 
When the double well is loaded with many atoms, 
we found that the decoherence process is slowed down by interactions and that the nature of the decay 
is altered and follows an algebraic law. 

The experimental realization of this model is possible using cold atoms trapped in optical 
lattices. One of the main challenges would be to use a system that can be well described by a single 
band model despite the large interaction strength. 
This could be achieved using a very light species like Lithium whose 
interaction can be tuned using a Feshbach resonance \cite{StreckerHulet2002, KhaykovichSalomon2002}. 
An alternate route would be to use two different kinds of bosonic atoms trapped in a single well potential 
and to control the effective tunneling and interaction respectively via two-photon combined microwave and
radio-frequency pulses, and a Feshbach resonance as in \cite{GrossOberthaler2010}. 

We thank P. Barmettler, H.P. Breuer, J. Dalibard, J.-P. Eckmann, M. Greiner, M. Lukin and 
V. Vuletic for fruitful discussions. We acknowledge ANR (FAMOUS), SNSF (Division II, MaNEP), 
CIFAR, NSERC of Canada and the DARPA-OLE program for financial support.

\bibliographystyle{prsty}

\begin{thebibliography}{10}
\bibitem{Hermitian} 
The other elements can be computed easily as the density matrix is Hermitian.
\end{thebibliography}

\end{document}